\PassOptionsToPackage{utf8}{inputenc}
\documentclass{bioinfo}
\copyrightyear{2021} \pubyear{2021}

\access{Advance Access Publication Date: 25 November 2021}
\appnotes{Applications Note}

\begin{document}
\firstpage{1}

\subtitle{Subject Section}

\title[DNA-Storage Errors Tool]{DNA Storage Error Simulator: A Tool for Simulating Errors in Synthesis, Storage, PCR and Sequencing}
\author[Alnasir \textit{et~al}.]{Thomas Heinis\,$^{\text{\sfb 1,}*}$, Jamie J. Alnasir\,$^{\text{\sfb 1,}*}$ and Louis Carteron\,$^{\text{\sfb 1,}}$}
\address{$^{\text{\sf 1}}$Department of Computing, Imperial College, London, SW7 2AZ, UK}
%\\
%$^{\text{\sf 2}}$Department, Institution, City, Post Code,
%Country.}
\corresp{$^\ast$To whom correspondence should be addressed.}

\history{}

\editor{}

\abstract{\textbf{Summary:} DNA has many valuable characteristics that make it suitable for a long-term storage medium, in particular its durability and high information density. DNA can be stored safely for hundreds of years with virtually no degradation, in contrast to hard disk drives which typically last for about 5 years. Furthermore, the duration of DNA-Storage can be extended to potentially up to thousands of years if it is desiccated and cooled in storage.\\
Advances in DNA technologies have made it possible to store the entirety of Wikipedia in a test tube and read that information using a handheld sequencing device, although imperfections in writing (synthesis) and reading (sequencing) need to be mitigated for it to be viable as a mainstream storage medium. New sequencing technologies, such as nanopore sequencing, aim to penetrate the consumer world, thanks to their affordability and size. However, the error characteristics of nanopore sequencing are not yet well characterised.\\
DNA Storage Error Simulator models errors that can be introduced in all the phases of DNA storage workflow, including synthesis, storage, PCR for amplification and finally sequencing. The error characteristics for sequencing and synthesis can be configured in all necessary detail or can be chosen from a predefined set of values based on available literature and our own analysis.\\
\textbf{Availability:} DNA Storage Error Simulator can be accessed online from
https://master.dbahb2jho41s4.amplifyapp.com/
(https://dnastorage.doc.ic.ac.uk/DNA-error-simulator)\\
\textbf{Contact:} \href{t.heinis@imperial.ac.uk}{t.heinis@imperial.ac.uk}, \href{j.alnasir@imperial.ac.uk}{j.alnasir@imperial.ac.uk}\\
%\textbf{Supplementary information:} Supplementary data are available at \textit{Bioinformatics} online
}

\maketitle

\section{Introduction}

%Figure~\ref{fig:01}

Advances in the encoding techniques used in DNA storage have helped push the information density of DNA towards its theoretical limit. Early works in the area achieved a low information density of 0.46 bits/nucleotides (b/nt) \citep{church2012next}, whereas the latest encoding techniques using fountain codes push this information density to 1.57 b/nt \citep{erlich2017dna}. One of the major limiting factors in achieving the theoretical limit of $\sim$2.0 b/nt is the necessity of incorporating error correction codes. Error correction codes are necessary to ensure the integrity of the data stored in DNA. Errors occur at all stages of the DNA storage channel: synthesis, storage, sequencing. Understanding the error properties of these three stages is essential to help facilitate the use of DNA as a commercial solution for data storage. Here we present DNA Storage Error Simulator, a web-based calculation tool for simulating errors in all stages of the DNA storage channel.

%\enlargethispage{12pt}

\section{DNA Storage Error Simulator}

The tool comprises parameter input panels for each phase of the DNA storage channel workflow, namely synthesis, storage, PCR, sequencing, which can be enabled or disabled.

\subsection{Synthesis}
The \textit{synthesis panel} of the calculator allows selection of the error characteristics --- substitution, insertion and deletion rate, respectively, as well as individual probabilities of substitution (i.e. A to T) --- either by using a dropdown box of predetermined dataset values or by manual input.

The datasets of pre-determined values provided are: Erlich, Goldman and PseudoRandom.

Errors in DNA synthesis are evenly distributed across a sequence. In constructing the output
sequence, it is feasible to decide at each nucleotide if an error should occur. The individual
error probabilities per nucleotide of substitutions, insertions and deletions are summed to give
the probability per nucleotide that an error has occurred. This number is used to decide during
the construction of the output if we should introduce an error using weighted coin method.

Even though insertions and substitutions are uniformly distributed across constructed sequences, deletions are tail favoured. DNA synthesis has a termination probability of 0.05\% per nucleotide.

Assuming an input length of 120 nt, the probability that the output is also 120 nt is only (1 - 0:05)\textsuperscript{120} $\approx 0.55 = 55\% $. Deletions can, therefore, be split into two different categories: true deletions and early termination deletions.

\textbf{Deletions:} Previous work on analysing errors during synthesis suggests that true deletions are non-discriminatory.
Any nucleotide is equally likely to be deleted at any point of the sequence. In addition to deletions, the termination factor of oligo synthesis is also accounted for during the construction of the synthesised output. A weighted coin approach determines whether the sequence should terminate early or should continue.

\textbf{Insertions:} Nucleotide insertions are not uniformly distributed. Insertions are dependent on the previous nucleotide in the sequence, due to the chemical binding nature of synthesis technologies. When the simulation dictates that an insertion should occur, the previous nucleotide is used as a key in a probability look-up table.

\textbf{Substitutions:} The outcome of a substitution is dependent on the nucleotide that is being replaced. Previous work
by \cite{heckel2019characterization} suggest that the transformations $C \rightarrow T$ (C becomes T) and $G \rightarrow A$ are the most common substitution types. A lookup table, constructed using cumulative probabilities, is used to determine what the outcome of the substitution should be. Previous work has shown that substitutions also depend on the previous nucleotides. However,
as the effect is minimal, to keep the model simple, only the current nucleotide is considered when choosing the replacement nucleotide.

\subsection{Storage}
The \textit{Storage panel} of the calculator is where the DNA storage temperature (in C) and period of storage (years) are input.

The decay of DNA is modelled by a standard radioactive decay. However, instead of removing nucleotides, the bonds between nucleotides are simply broken. This means that the broken strands are no longer readable as the forward and reverse primer are no longer located on the same strand.

\begin{equation}
\lambda = 41.2 - 15267.6 \frac{1}{T}\label{eq:01}\vspace*{-10pt}
\end{equation}\\

The expected half-life t\textsubscript{1/2} of DNA can be calculated by the following equation:

\begin{equation}
t_{1/2}=\frac{ln 2}{\lambda}\label{eq:01}\vspace*{-10pt}
\end{equation}

\subsection{PCR}
The \textit{PCR panel} provides input for the number of PCR cycles.

PCR is one of the more well-characterised phases of the DNA storage channel. It is usually used before sequencing (except nanopore) to amplify certain sequences through primers, resulting in an overwhelming representation of that specific sequence in the final pool. PCR is often modelled as a standard exponential relation, namely by:

\begin{equation}
T_{n+1} = T_{n}*(1+E_{n}); E_{n} \in (0,1)
\end{equation}

$T_{n}$ is the product yield and $E_{n}$ is the amplification efficiency. A higher amplification efficiency means more yield.
Work by \cite{alvarez2007model} show that it would be better to model the amplification efficiency as a sigmoid function of the product yield. Namely, they model $E_{n}$ based on Tn as follows:

\begin{equation}
E_{n} = (1+e^{\frac{T_{n}-T_{m}}{b}})
\end{equation}

$T_{m}$ and b are parameters to be fitted by non-linear regression. This model helps dictate the total concentration of sequences after n PCR cycles.

\subsection{Sequencing}
The \textit{sequencing panel}, appearing similar to that of the \textit{synthesis panel}, allows selection of the error characteristics of the sequencing process. The approach to modelling sequencing is very similar to modelling the synthesis stage --- the
probabilities for substitution, deletions and insertions, in addition to their respective lookup tables,
have been changed to achieve this.

\cite{heckel2019characterization} state that the most common error type during sequencing is substitutions. The
substitution reading error rates are reported to be about 0.0015 - 0.0004 errors per base for Illumina
sequencing. Insertions and deletions happen a lot less frequently, in the order of 10\textsuperscript{6}. As substitutions are the most common error, they can be easily mitigated using common error-correcting algorithms.

\section{Conclusions}

DNA Storage Error Simulator simulates errors at all stages of the DNA storage workflow --- in sequencing, storage, PCR and sequencing. The simulation is based on data obtained from relevant literature, and for the synthesis and sequencing phases, pre-determined probabilities for insertions, deletions and probabilities can be used from pre-existing datasets. The tool can also simulate errors in PCR, given the number of cycles, as well as degradation of DNA in the storage phase given the number of years and temperature the DNA was stored at. DNA Storage Error Simulator tool can be used to determine the level of redundancy to be employed in error correction for a given DNA storage workflow.

\section*{Acknowledgements}

We thank HelixWorks who kindly provided the data sets used in developing this tool.

\vspace*{-12pt}

\section*{Funding}

This work was partially funded by the European Union’s Horizon 2020 research and innovation programme, project OligoArchive (grant agreement No 863320).

\bibliographystyle{natbib}
\bibliography{document}

% \begin{thebibliography}{}

% \bibitem[Bofelli {\it et~al}., 2000]{Boffelli03}
% Bofelli,F., Name2, Name3 (2003) Article title, {\it Journal Name}, {\bf 199}, 133-154.

% \bibitem[Bag {\it et~al}., 2001]{Bag01}
% Bag,M., Name2, Name3 (2001) Article title, {\it Journal Name}, {\bf 99}, 33-54.

% \bibitem[Yoo \textit{et~al}., 2003]{Yoo03}
% Yoo,M.S. \textit{et~al}. (2003) Oxidative stress regulated genes
% in nigral dopaminergic neurnol cell: correlation with the known
% pathology in Parkinson's disease. \textit{Brain Res. Mol. Brain
% Res.}, \textbf{110}(Suppl. 1), 76--84.

% \bibitem[Lehmann, 1986]{Leh86}
% Lehmann,E.L. (1986) Chapter title. \textit{Book Title}. Vol.~1, 2nd edn. Springer-Verlag, New York.

% \bibitem[Crenshaw and Jones, 2003]{Cre03}
% Crenshaw, B.,III, and Jones, W.B.,Jr (2003) The future of clinical
% cancer management: one tumor, one chip. \textit{Bioinformatics},
% doi:10.1093/bioinformatics/btn000.

% \bibitem[Auhtor \textit{et~al}. (2000)]{Aut00}
% Auhtor,A.B. \textit{et~al}. (2000) Chapter title. In Smith, A.C.
% (ed.), \textit{Book Title}, 2nd edn. Publisher, Location, Vol. 1, pp.
% ???--???.

% \bibitem[Bardet, 1920]{Bar20}
% Bardet, G. (1920) Sur un syndrome d'obesite infantile avec
% polydactylie et retinite pigmentaire (contribution a l'etude des
% formes cliniques de l'obesite hypophysaire). PhD Thesis, name of
% institution, Paris, France.

% \end{thebibliography}
\end{document}